\documentclass[twocolumn]{aastex62}

\newcommand{\alw}{a$_{<LW>}$}
\newcommand{\amw}{a$_{<MW>}$}

\newcommand{\mfit}{M$_{*,spec}$}
\newcommand{\lfit}{L$_{spec}$}

\newcommand{\mdot}{M$_{\odot}$}

\accepted{April 5, 2019}

\submitjournal{ApJ}

\shorttitle{REJUVENATION IN QUIESCENT GALAXIES}
\shortauthors{Chauke et al.}

\begin{document}

\title{REJUVENATION IN $z\sim0.8$ QUIESCENT GALAXIES IN LEGA-C}

\email{chauke@mpia-hd.mpg.de}

\author[0000-0002-0786-7307]{Priscilla Chauke}
\affil{Max-Planck-Institut f\"ur Astronomie, K\"onigstuhl 17, D-69117 Heidelberg, Germany}

\author{Arjen van der Wel}
\affiliation{Sterrenkundig Observatorium, Universiteit Gent, Krijgslaan 281 S9, B-9000 Gent, Belgium}

\author{Camilla Pacifici}
\affiliation{Space Telescope Science Institute, 3700 San Martin Drive, Baltimore, MD 21218, USA}

\author{Rachel Bezanson}
\affiliation{University of Pittsburgh, Department of Physics and Astronomy, 100 Allen Hall, 3941 O'Hara St, Pittsburgh PA 15260, USA}

\author{Po-Feng Wu}
\affiliation{Max-Planck-Institut f\"ur Astronomie, K\"onigstuhl 17, D-69117 Heidelberg, Germany}

\author{Anna Gallazzi}
\affiliation{INAF-Osservatorio Astrofisico di Arcetri, Largo Enrico, Fermi 5, I-50125 Firenze, Italy}

\author{Caroline Straatman}
\affiliation{Sterrenkundig Observatorium, Universiteit Gent, Krijgslaan 281 S9, B-9000 Gent, Belgium}

\author{Marijn Franx}
\affiliation{Leiden Observatory, Leiden University, PO Box 9513, 2300 RA Leiden, The Netherlands}

\author{Ivana Bari\v{s}i\'{c}}
\affiliation{Max-Planck-Institut f\"ur Astronomie, K\"onigstuhl 17, D-69117 Heidelberg, Germany}

\author{Eric F. Bell}
\affiliation{Department of Astronomy, University of Michigan, 1085 S. University Ave., Ann Arbor, MI 48109, USA}

\author{Josha van Houdt}
\affiliation{Max-Planck-Institut f\"ur Astronomie, K\"onigstuhl 17, D-69117 Heidelberg, Germany}

\author{Michael V. Maseda}
\affiliation{Leiden Observatory, Leiden University, PO Box 9513, 2300 RA Leiden, The Netherlands}

\author{Adam Muzzin}
\affiliation{Department of Physics and Astronomy, York University, 4700 Keele St., Toronto, Ontario, MJ3 1P3, Canada}

\author{David Sobral}
\affiliation{Leiden Observatory, Leiden University, PO Box 9513, 2300 RA Leiden, The Netherlands}
\affiliation{Physics Department, Lancaster University, Lancaster LA1 4YB, UK}

\author{Justin Spilker}
\affiliation{Department of Astronomy, University of Texas at Austin, 2515 Speedway, Stop C1400, Austin, TX 78712, USA}




\begin{abstract}

We use reconstructed star-formation histories (SFHs) of quiescent galaxies at $z=0.6-1$ in the LEGA-C survey to identify secondary star-formation episodes that, after an initial period of quiescence, moved the galaxies back to the star-forming main sequence (blue cloud). $16\pm3$\% of the $z\sim0.8$ quiescent population has experienced such rejuvenation events in the redshift range $0.7<z<1.5$ after reaching quiescence at some earlier time. On average, these galaxies first became quiescent at $z=1.2$, and those that rejuvenated, remained quiescent for $\sim1$Gyr before their secondary SF episode which lasted $\sim0.7$Gyr. The stellar mass attributed to rejuvenation is on average 10\% of the galaxy stellar mass, with rare instances of an increase of more than a factor 2. Overall, rejuvenation events only contribute $\sim2$\% of the total stellar mass in $z\sim0.8$ quiescent galaxies and we conclude that rejuvenation is not an important evolutionary channel when considering the growth of the red sequence. However, our results complicate the interpretation of galaxy demographics in color space: the galaxies with rejuvenation events tend to lie in the so-called `green valley', yet their progenitors were quiescent at $z\sim2$.

\end{abstract}




\section{Introduction}
\label{sec:intro}

The colors of galaxies are known to be bimodal, not only in the local universe \citep[e.g.,][]{strateva2001,baldry2004}, but also at redshift $z\sim1$ and beyond \citep[e.g.][]{bell2004,franzetti2007,whitaker2011,straatman2016}. Galaxies are classified as either part of the `blue cloud' or `red sequence', where the blue cloud contains galaxies that are actively forming new stars, while the red sequence contains quiescent galaxies that have very low ongoing star-formation (SF). Ages and metallicities of massive quiescent galaxies (stellar mass $\gtrsim10^{10.5}M_{\odot}$) at $z\sim1$ are consistent with passive evolution to the present-day \citep[e.g.,][]{gallazzi2014,choi2014}, although \citet{gallazzi2014} require additional quenching of a fraction of massive star-forming galaxies at $z\sim1$ to account for the scatter in the ages of present-day quiescent galaxies. \citet{schiavon2006}, however, compared stacked spectra of red sequence galaxies at redshifts $0.7\leq z \leq1$ to local SDSS galaxies \citep{york2000} and found that their ages are inconsistent with passive evolution, which suggests that either new galaxies with younger stars continually transition to the red sequence, or individual quiescent galaxies experience `frosting', where continuing low-level star formation adds a minority of young stars to an older base population \citep{trager2000}. \citet{wu2018} and \citet{spilker2018} reached the same conclusion using high-resolution spectra of massive (stellar mass $>10^{11}$\mdot) $z\sim0.8$ galaxies from the Large Early Galaxy Astrophysics Census Survey \citep[LEGA-C,][]{vdw2016}.

The increasingly dominant population of quiescent galaxies measured in number density evolution studies \citep[e.g.,][]{pozzetti2010, bram2011, moustakas2013, muzzin2013g} indicates that star-forming galaxies have their star-formation quenched and transition from the blue cloud to the red sequence. The nature of this quenching process is still not understood, although the `maintenance mode' of AGN feedback is widely believed to suppress star-formation in massive galaxies by providing sufficient energy to keep the halo gas from cooling \citep[e.g.,][]{goto2006,heckman2014}. Therefore, if AGN feedback fails to keep halo gas hot, the star-formation in a galaxy could be reignited. 

Although the aforementioned evolution studies indicate that galaxies evolve from being star-forming to quiescent, secondary SF has been found to be a common phenomenon in local early-type galaxies: the fraction of early-type galaxies showing evidence of recent star formation is thought to be between $\sim$10 and 30\% \citep{schawinski2007,donas2007}. This fraction is higher in low-density environments \citep[e.g.,][]{schawinski2007,thomas2010}, which is consistent with {H I} being detected more often in field galaxies than in clusters \citep[e.g.,][]{osterloo2010}. \citet{treu2005} and \citet{thomas2010} found that the fraction of stellar mass formed from secondary SF episodes decreases with galactic mass, ranging from $<1$\% for stellar masses $>10^{11.5}$\mdot\ to $\sim10\%-40\%$ for stellar masses $<10^{11}$\mdot. This is in line with \citet{kaviraj2007} who found that star-formation is more efficiently quenched in high-mass galaxies (for stellar masses $>10^{10}$\mdot).

Secondary SF episodes have been linked to either {H I} gas accretion or mergers which bring in gas, often resulting in only a small population of relatively young stars \cite[e.g.][]{yi2005,kaviraj2009,marino2009}. There is no general trend between stellar population and {H I} properties, however, galaxies with a significant young sub-population have inner gas discs \citep{osterloo2010}.  Post-starburst (PSB) or `E+A' galaxies, i.e. young quiescent galaxies with strong Balmer absorption lines and weak to no SF-related emission lines \citep{dressler1983,dressler1999,tran2004}, have also been linked to secondary SF episodes in order to reconcile the number density of PSBs with the slow growth of the quiescent population at the high-mass end ($>10^{11}$\mdot) at $z<1$ \citep[e.g.,][]{rowlands2018}, as well as using starburst timescales ($\sim500$Myr) to show that PSBs are likely not a major component in the growth of the passive galaxy population \citep[e.g.,][]{dressler2013}. E+A galaxies are also likely caused by interactions or mergers \citep{goto2005,yamauchi2008}.

There are multiple measurements of the fraction of galaxies that undergo secondary SF, however, previous studies were mostly limited to low redshifts because of the abundance of high-resolution spectra in the local Universe. Furthermore, many of these studies do not determine whether the galaxies `rejuvenate', i.e. transition back to the blue cloud from the red sequence. Recently, \citet{pandya2017} analysed a semi-analytical model of galaxy formation as well as GAMA and CANDELS observations out to $z=3$ to constrain the frequency of rejuvenation episodes. They measured the transition of massive galaxies ($>10^{10}$\mdot) from the star-forming `main sequence' (SFMS) to both the `transition region' between the blue cloud and red sequence, and the quiescent sequence. \added{Using their semi-analytical model,} they found that the average $z=0$ quiescent galaxy first joined the quiescent population at $z\sim0.4$ and that 31\% of quiescent galaxies have experienced at least one rejuvenation event since $z=3$. However, these rejuvenation timescales are short as the average time a galaxy spends in quiescence \added{between $z=3$ and $z=0$} is comparable \replaced{to that of an average galaxy that does not undergo rejuvenation}{for rejuvenated and non-rejuvenated galaxies in their model. \citet{behroozi2019} applied empirical models of galaxy formation to dark halo merger trees to determine individual galaxies' SFRs that are consistent with observations (e.g. stellar mass functions, specific and cosmic SFRs, quenched fractions, etc.). They found that, at $z=1$, rejuvenation fractions range from $\sim10\%$ to $\sim20$\% for stellar masses in the range $10^{10}-10^{11}$\mdot, with the rejuvenation fraction peaking around $4\times10^{10}$\mdot. Their rejuvenation fractions at $z=0$ are significantly higher ($\sim30-60$\% in the same stellar mass range), presumably because galaxies at lower redshifts have had more time to quench and then rejuvenate at a later stage.}


\begin{figure*}[t]
\centering
\includegraphics[width=1.\textwidth]{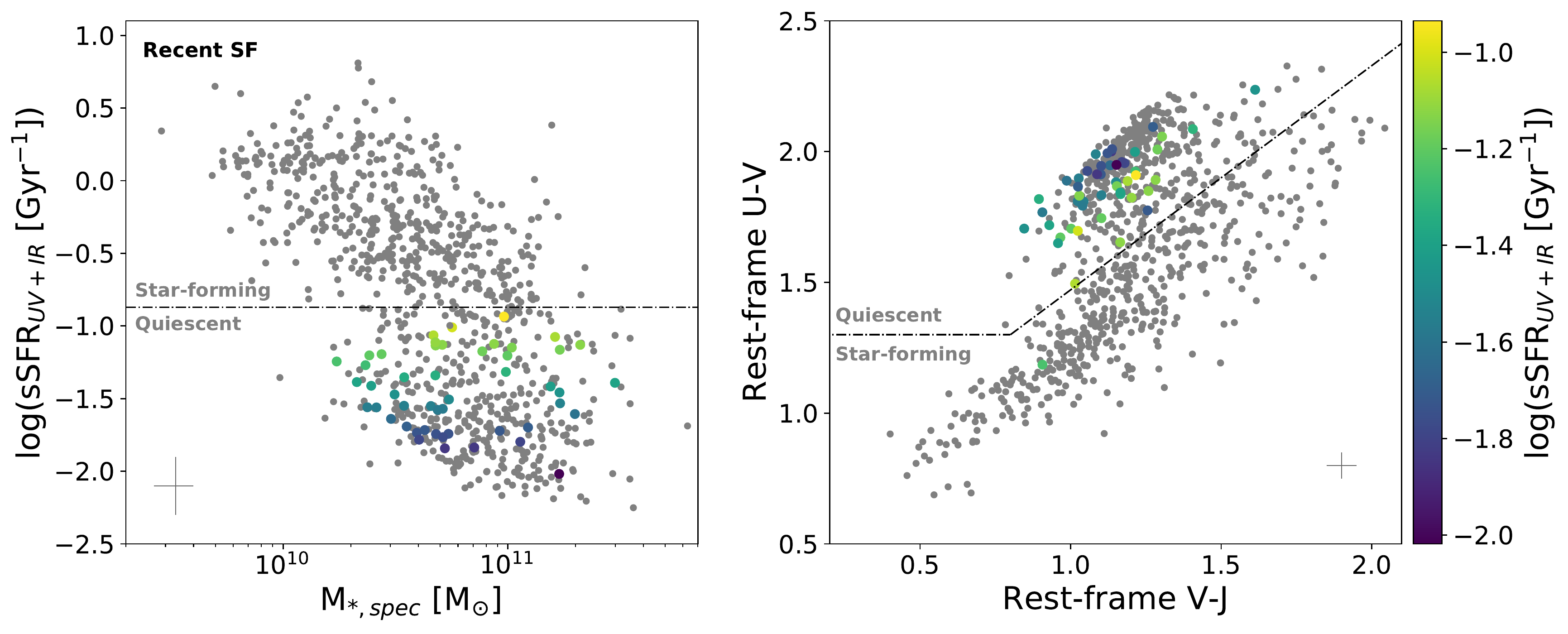}
\caption{sSFR$_{UV+IR}$ as a function of \mfit\ (left) and the rest-frame UVJ diagram (right) of the LEGA-C population. The dashed line distinguishes the star-forming and quiescent populations. The rejuvenated quiescent population is color-coded by sSFR$_{UV+IR}$ for comparison. Typical error bars are indicated in dark gray.}
\label{fig:uvj}
\end{figure*}

The magnitude of the effect of rejuvenation processes still needs to be addressed. Specifically, we do not know if the stellar mass formed from such events is a significant portion of the cosmic star-formation rate density (SFRD) of the Universe and on what timescale these events occur, or how often secondary SF episodes cause galaxies to transition back to the blue cloud during the event \added{(i.e. rejuvenation)}. In an earlier paper \citep{chauke2018}, we reconstructed the star-formation histories (SFHs) of galaxies at $z\sim1$ using the high-resolution spectra from LEGA-C. These SFHs revealed secondary star-formation episodes in a minority of the quiescent population. This allows us to be able to investigate rejuvenation timescales as well as the frequency and magnitude of such events. In this study, we use the reconstructed SFHs from \citet{chauke2018} to investigate quiescent galaxies with \replaced{secondary SF}{rejuvenation} episodes. In Section \ref{sec:data} we give a brief overview of the sample. In Section \ref{sec:disc} we investigate the properties of rejuvenated galaxies, viz. the timescales of \replaced{secondary SF}{rejuvenation} episodes, the \added{local environmental} density and mass dependence of these episodes, as well as whether the mass formed from such events is significant. Finally, in Section \ref{sec:summ} we summarise the results. We assume a $\Lambda{CDM}$ model with $H_0=67.7$km\,s$^{-1}$Mpc$^{-1}$, $\Omega_m=0.3$ and $\Omega_{\Lambda}=0.7$.

\section{Data}
\label{sec:data}

LEGA-C \citep{vdw2016} is an ESO Public Spectroscopic survey with VLT/VIMOS of $\sim$3000 galaxies in the COSMOS field with redshifts in the range $0.6<z<1.0$. The galaxies were selected from the Ultra-VISTA catalog \citep{muzzin2013f}, with a redshift dependent K-band limit (K$_{ab} = 20.7-7.5\times$log$[(1+z)/1.8]$). Each galaxy is observed for $\sim20$\,h, which results in spectra with $S/N\sim20$\AA$^{-1}$ (with resolution R\,$\sim3000$) in the wavelength range $\sim0.6\mu{m}-0.9\mu{m}$. For details of the data reduction procedure, see \cite{vdw2016} and \cite{straatman2018}. This work is based on the second data release\footnote{\url{http://www.eso.org/sci/publications/announcements/sciann17120.html}}, which contains 1550 primary sample galaxies. We make use of the following measured quantities in the analysis: rest-frame U-V and V-J colors, UV+IR star formation rates (SFRs), \added{stellar masses (\mfit),} UV+IR specific SFRs (sSFR$_{UV+IR}$\added{, i.e. UV+IR SFRs divided by \mfit}), and scale-independent local \replaced{densities}{overdensities, log$(1+\delta)$, i.e. the local surface density divided by the mean local surface density.} The UV+IR SFRs are estimated from UV and IR luminosities, following \cite{whitaker2012}. \added{{\mfit} is the stellar mass estimate obtained from full-spectrum fitting (see Section \ref{sec:datsfh}).} The \replaced{scale-independent local densities}{log$(1+\delta)$ values} are estimated from redshift slices using the Voronoi tessellation method \citep{darvish2016}. Figure \ref{fig:uvj} shows sSFR$_{UV+IR}$ as a function of \replaced{stellar mass \mfit\ (left panel, see Section \ref{sec:datsfh})}{\mfit\, (left panel),} and the rest-frame UVJ diagram (right panel) of the LEGA-C population at the observed redshift. Star-forming and quiescent populations are distinguished by the dashed lines and quiescent galaxies whose SF was rejuvenated, are color-coded by sSFR$_{UV+IR}$ for comparison. See Section \ref{sec:datsamp} for definitions of quiescence and rejuvenation.

\subsection{Star Formation Histories}
\label{sec:datsfh}

\citet{chauke2018} used a custom full-spectrum fitting algorithm to reconstruct the SFHs of the LEGA-C sample. The algorithm incorporates \textit{emcee} \citep[an affine invariant ensemble sampler for MCMC, ][]{foreman2013} and \textit{FSPS v3.0} \citep[the Python implementation of the Flexible Stellar Population Synthesis package,][]{conroy2010, conroy2009, foreman2014}. \textit{emcee} makes use of MCMC `walkers' which randomly explore the parameter space and converge to the most likely parameters values. The galaxy spectra were fit to a linear combination of a set of 12 composite stellar populations (CSPs), with solar metallicity and constant star formation within the time interval of the templates. The algorithm \added{uses \citet{calzetti2000}'s dust reddening curve to fit}\deleted{also fits} for 2 dust reddening values, $E(B-V)_i$, one for the youngest (more dust-obscured) stellar population \added{($0-100$Myr)} and the second for the other stellar populations. The algorithm uses emission-line subtracted spectra. Emission line spectra are computed using the Penalized Pixel-Fitting method \cite[pPXF, ][]{cappellari2004}. For details of the emission line fitting procedure, see \cite{bezanson2018a}. The model results in measurements of stellar masses (\mfit), luminosities (\lfit), mean mass-weighted and light-weighted ages (\amw\ and \alw, respectively) and the dust reddening values. The stellar masses derived using our method, \mfit, are in good agreement with photometry-based stellar masses derived with FAST \citep{kriek2009}. See \citet{chauke2018} for further details and results of the fitting algorithm.

\begin{figure*}[th]
\centering
\epsscale{0.9}
\plotone{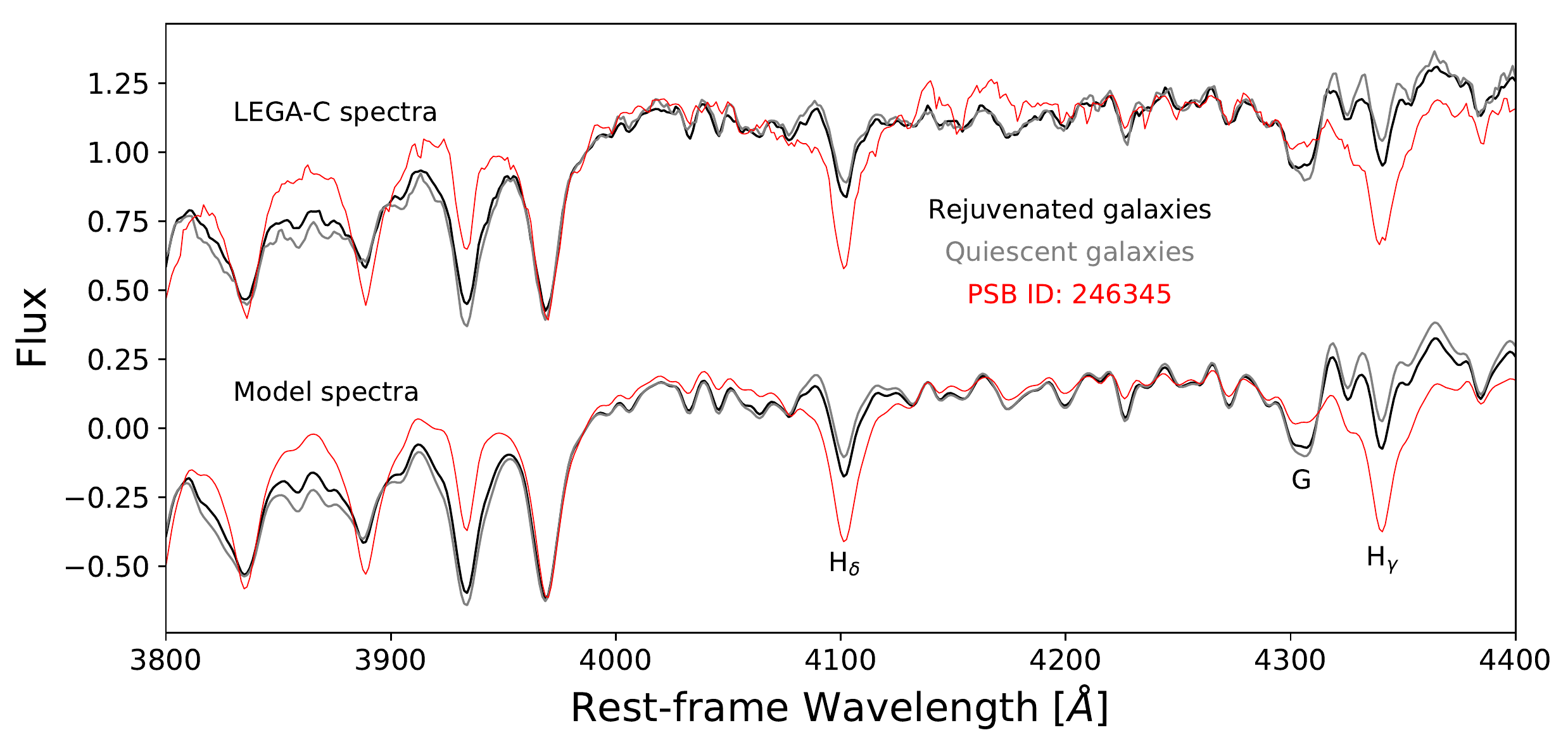}
\caption{Average spectrum (LEGA-C as well as best-fit model) of rejuvenated galaxies (black) compared to the average spectrum of stellar mass and H$_\delta$ matched  quiescent galaxies that do not show evidence of rejuvenation (gray). The PSB spectrum is shown for comparison. The spectra have been normalised and shifted for comparison purposes.}
\label{fig:specav}
\end{figure*}

\subsection{Identifying Rejuvenated Galaxies}
\label{sec:datsamp}

\added{In this section we identify a sample of galaxies that, according to our spectral fits, have a large probability of having experienced a rejuvenation event after an initial period of quiescence.} One of the main goals of this study is to measure the contribution of rejuvenation to the total stellar mass of quiescent galaxies, therefore, we select quiescent galaxies by their specific star-formation rate (sSFR). This approach results in an identical sample of rejuvenated galaxies, with the exception of one galaxy, compared to selecting quiescent galaxies by their U-V and V-J colors, see Figure \ref{fig:uvj} for a comparison at the observed redshift. In this study, sSFR is used as a quiescent selection criterion because we can directly compute the sSFR from our reconstructed SFHs. 

We use the reconstructed SFHs to compute the sSFR of a galaxy at time $t_i$, where $i$ is the time bin number and sSFR is the SFR at t$_i$ divided by the total stellar mass at that time, i.e. $sSFR_{t_i} = SFR_{t_i}/M_{*,t_i}$. A galaxy is defined as quiescent if the log(sSFR [Gyr$^{-1}]) <-1$ at redshift $z=0.8$. To track the quiescence of galaxies along their SFHs, we use the \citet{speagle2014} redshift dependent sSFR to adjust our definition for quiescence to larger redshifts. For example, the limit increases to $-0.40$ and $-0.03$ at redshifts $z=2$ and $z=3$, respectively \citep[see ][for full details]{speagle2014}. The instant a galaxy's sSFR falls below the redshift adjusted limit at some point in its history, it is considered to be quiescent at that point (and the accumulated stellar mass is recorded), until such a time that it rises above the limit. The stellar mass accumulated during this period is considered to be mass formed from secondary SF. 

To minimise the error of adding false positives, we compute the probability of a rejuvenation event having occurred using the MCMC walkers (see Section \ref{sec:datsfh} and Figure \ref{fig:fit3}). The probability that a galaxy has a rejuvenation episode ($p_{REJ}$) is equal to the probability that the galaxy is quiescent at $t_i$ ($p_{Q_i}$) multiplied by the probability that it is star-forming at a later time $t_j$ ($p_{{SF}_{j}}$). We find the time interval where the galaxy is quiescent (i.e. $p_{Q_{i}}>0.5$) and the interval where it is star-forming, and then compute the maxima of the product of the probabilities over these time intervals. $p_{Q_i}$ (\replaced{$p_{{SF}_{i+1}}$}{$p_{{SF}_{j}}$}) is the fraction of walkers that lie below (above)  the sSFR limit at $t_i$ (\replaced{$t_{i+1}$}{$t_{j}$}). If $p_{REJ}>0.5$, then the galaxy is considered to have had a rejuvenation episode. \added{For example, 91529 (see Figure \ref{fig:fit3}) is star-forming in bin 2 then has a quiescent period in bin 3 and 4, which is followed by a period of rejuvenation from bin 5 to 8 before the galaxy is quiescent again from bin 9 to 12. In this case, $p_{REJ} = p_{Q_3}\times{p_{SF_5}} = 1$, and the galaxy is classified as a rejuvenated galaxy. If a galaxy alternates between quiescent and star-forming multiple times during its SFH, then $p_{REJ}$ is calculated as above for each episode. If the $p_{REJ}$ requirement is satisfied, then the episode is counted as a rejuvenation episode. We also compute how long the galaxy was quiescent for before the first rejuvenation episode, i.e. the time range where $p_{Q}>0.5$ after the initial SF episode and before rejuvenation (bin 3 to bin 4 for 91529), as well as measure how long the rejuvenation episode lasts, i.e. the time range where $p_{SF}>0.5$ during the rejuvenation episode (bin 5 to 8 for 91529).}

All galaxies in our sample, with the exception of 3 \added{(viz. 206858, 228340 and 92132)}, contain at most 1 rejuvenation episode before becoming quiescent again at the observed redshift. The quiescent sample contains 412 galaxies, 52 ($13$\%) of which have had a rejuvenation episode. These galaxies are shown in Figure \ref{fig:uvj}, they are color-coded by sSFR$_{UV+IR}$ for comparison with the rest-frame UVJ diagram on the right panel. If the sSFR limit were decreased (increased) to $-1.3$ ($-0.7$), the fraction of rejuvenated galaxies in our sample would be 19\% (9\%). Furthermore, using a combination of sub-solar ($0.4$\,Z$_\odot$) and solar metallicity CSPs as well as super-solar ($2.5$\,Z$_\odot$) and solar metallicity CSPs in our fitting algorithm, instead of solar metallicity CSPs alone, results in \deleted{the sample of rejuvenated galaxies changing by 24 and 26\%, respectively.} \added{76 and 74\% of the sample, respectively, remaining the same. Therefore, mixing metallicities does not significantly change the resulting sample; however, it is unclear how exactly metallicity errors affect the rejuvenation fraction because of the age-metallicity degeneracy in stellar populations. Both the rejuvenated and non-rejuvenated quiescent populations have S/N $\sim20$, characteristic of LEGA-C galaxies. We have shown in \citet{chauke2018}, using noisy synthetic spectra, that our fitting algorithm converges for this S/N.}

In Figure \ref{fig:specav}, we compare the average normalised spectrum (LEGA-C as well as best-fit model) of quiescent galaxies that are identified as rejuvenated to stellar mass and H$_\delta$ matched quiescent galaxies \added{(i.e. within $0.01$dex for stellar mass and $0.5${\AA} for H$_\delta$)} that are not identified as such. A PSB spectrum is also shown for comparison. The average rejuvenated galaxy has stronger $H_\delta$ and $H_\gamma$ lines, which are characteristic of young stellar populations; however, its G-band (absorption lines of the CH molecule around 4300{\AA}), which is characteristic of older stellar populations, is not as strong as that of quiescent galaxies without rejuvenation episodes. See Figure \ref{fig:fit1} in the appendix for individual best-fit spectra of rejuvenated galaxies in our sample obtained from MCMC full-spectrum fitting \citep{chauke2018}. Compared to PSB or E+A galaxies, our sample of rejuvenated galaxies have weaker Balmer lines and redder \added{V-J} colors. Their V-J colors (see Figure \ref{fig:uvj}) are higher than the typical cut for PSBs \citep[V-J $\lesssim1$, e.g.][]{whitaker2012}, and most have H$\delta$ equivalent widths (EW[H$\delta$], see Figure \ref{fig:specav}) that are lower than required for PSBs \citep[EW[H$\delta{]}\sim3-5$,][]{wu2018}. Therefore, our sample of rejuvenated galaxies \replaced{indicates}{suggests} that most PSB and E+A galaxies are not recently rejuvenated galaxies as suggested in previous studies \citep[e.g.,][]{dressler2013,rowlands2018}; instead, they \replaced{are}{might be} galaxies that recently quenched for the first time.

\added{
\subsection{Determining the Rejuvenation Fraction}
\label{sec:walk}}

\added{In this section we consider, for each individual galaxy, the probability that a rejuvenation event occurred and use the sum of those probabilities to assess the importance of rejuvenation events in the context of the cosmological SFH. Some rejuvenation events will be missed by the selection criteria described in Section \ref{sec:datsamp} above, which motivates us to consider the probability that a rejuvenation event occurred in each quiescent galaxy. For this purpose, we use the MCMC walkers to compute the probability that a galaxy is quiescent ($p_{Q_i}$) or star-forming ($p_{{SF}_{i}}$) for each of the 12 time bins $i$. We find the time interval where the galaxy is quiescent (i.e. $p_{Q_{i}}>0.5$) after an initial period of SF, and compute the maximum in this range ($p_{Q_{qmax}}$), i.e. when the galaxy has the highest chance of being quiescent. $p_{Q_{qmax}}$ is multiplied to $p_{{SF}_{i}}$, where $i$ ranges from $qmax+1$ to 12 (the youngest bin), and then the maximum of $p_{Q_{qmax}}\cdot{}p_{{SF}_{i}}$ is determined. Finally, we compute the sum of these maxima for all galaxies in the quiescent sample to determine the fraction of galaxies with rejuvenation episodes. We also measure the mass formed between the first time a galaxy reached quiescence ($i=q1$) and $i=12$.}

\added{We find that $16\pm3$\% of the quiescent population has returned to the star-forming sequence during the epoch $z\sim0.7-1.5$ after reaching quiescence at some earlier time. This is consistent with the fraction of galaxies that have been identified as rejuvenated (Section \ref{sec:datsamp}).  However, these rejuvenation events account for only $2\pm1$\% of the stellar mass in quiescent galaxies at $z\sim0.8$. We have applied a completeness correction to these measurements on a galaxy-by-galaxy basis as described by \citet{wu2018} to create a volume-limited quantity. The uncertainties are estimated by bootstrapping the sample. These numbers are based on the requirement that a galaxy moves back to the SF sequence, as described in Sec \ref{sec:datsamp}.  If we repeat our calculation with a less strict requirement, namely that SF exceeds a fixed value of  log(sSFR [Gyr$^{-1}]) <-1$ at any redshift after initial quiescence, instead of an evolving limit, the fraction of galaxies with rejuvenation events increases to $24\pm2$\%, with a total mass contribution of $4\pm1$\%.} 

\added{Our rejuvenation fraction lies between \citet{donas2007} and \citet{schawinski2007}'s measured values of 10\% and 30\%, respectively, for $z\sim0.1$ galaxies, though we note that our value is a lower limit  as we only trace rejuvenation episodes in the redshift range $0.7<z<1.5$ due to limitations in the reconstructed SFHs (see Section \ref{sec:time}). Furthermore, \citet{donas2007} and \citet{schawinski2007} use UV-color relations to trace recent SF, therefore, their methods trace secondary SF that does not necessarily lead galaxies back to the the star-forming sequence.} 

\added{\citet{pandya2017} also measure a higher rejuvenation fraction (31\%) in their SAM (see Section \ref{sec:intro}). They define the quiescent region to be $1.4$dex below the SFMS, considerably more strict than our definition, and they additionally define a transition region ($0.6-1.4$dex below SFMS) between the star-forming and quiescent sequence. Defining our quiescent region in the same manner results in our rejuvenation fraction decreasing to 5\%. However, if we instead define the quiescent region to be $0.6$dex below the SFMS, i.e. combine \citet{pandya2017}'s transition and quiescent population, our rejuvenation fraction increases to 18\%, still lower than their value of 31\%. The discrepancy between these results may suggest that our sample contains galaxies with hidden rejuvenation at larger lookback times ($\gtrsim10$Gyrs), or that the fraction of rejuvenated galaxies is higher at $z=0$. However, we note that \citet{pandya2017}'s SAM underproduces quiescent galaxies at $z>0.5$, and they suggest that one of the reasons could be that quiescent galaxies are rejuvenating too much in their SAM.}

\added{Our rejuvenation fraction is in agreement with \citet{behroozi2019}'s measurements ($\sim10-20\%$). Their SFR distribution is assumed to be the sum of two log-normal distributions corresponding to a quenched population and a star-forming population, at fixed redshift and peak circular velocity at the redshift of peak halo mass, and rejuvenation is defined as at least 300 Myr of quiescence followed by at least 300 Myr of star formation.}

\begin{figure}[t]
\centering
\gridline{\fig{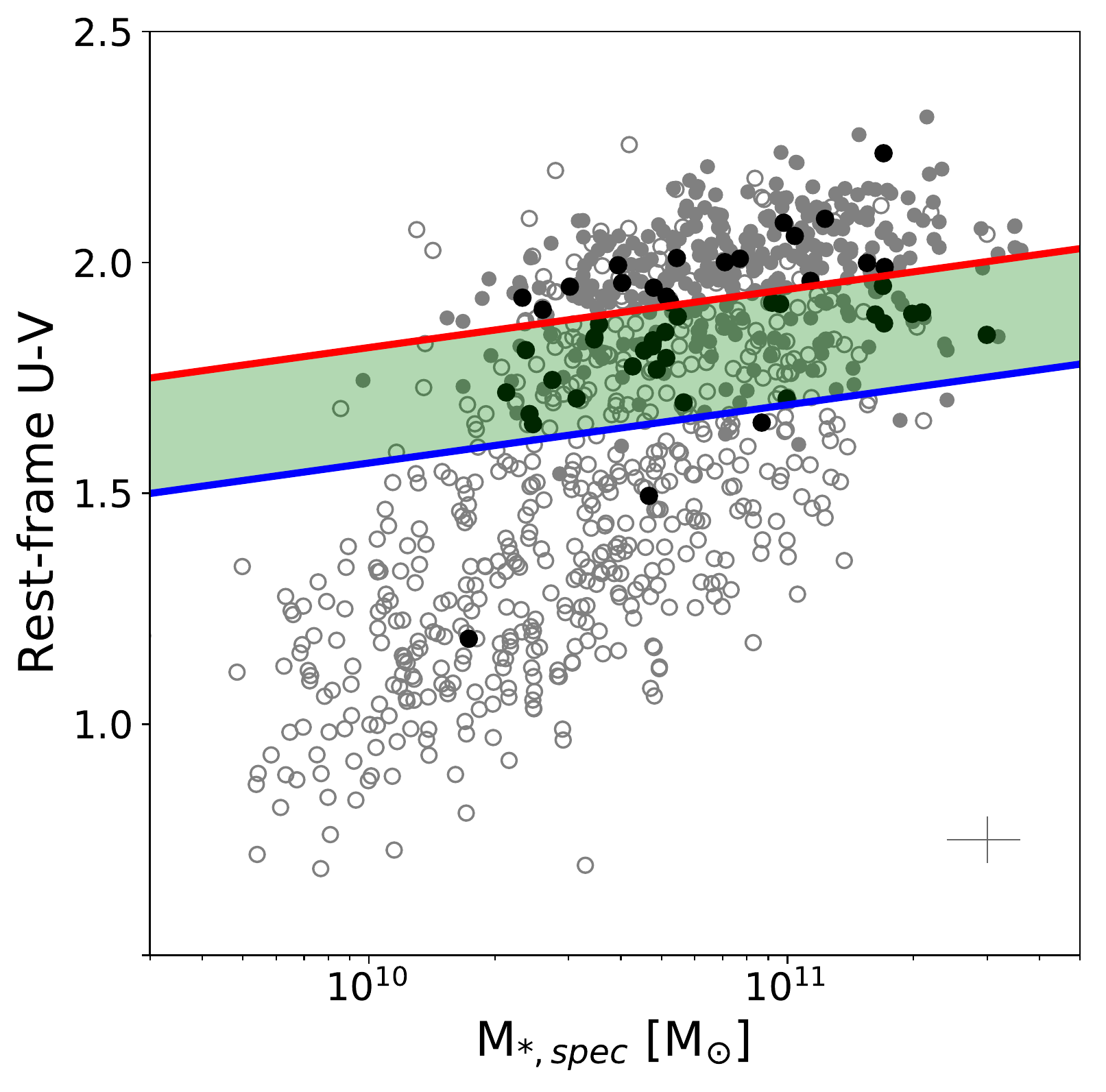}{0.45\textwidth}{}}
\caption{Rest-frame U-V color versus \mfit\ of the quiescent (filled gray circles) and star-forming (open gray circles) populations in the LEGA-C sample. The black points represent rejuvenated quiescent galaxies, while the green band and the blue and red lines indicate the green valley, star-forming and quiescent regions, respectively. Typical error bars are indicated in dark gray.}
\label{fig:uvm}
\end{figure}

\section{Properties of Rejuvenated population}
\label{sec:disc}

\subsection{The Green Valley}
\label{sec:gv}

Figure \ref{fig:uvm} shows the rest-frame U-V color as a function of \mfit\ of rejuvenated galaxies compared to the LEGA-C sample as a whole. The majority of galaxies with rejuvenation episodes have intermediate U-V colors and stellar masses (as well as sSFRs, see Figure \ref{fig:uvj}), i.e. they are in the so-called `green-valley', where galaxies are thought to be in the transition phase from the blue cloud to the red sequence. This is to be expected since recent SF should boost the U-V color and rejuvenation episodes decrease with stellar mass  \citep{treu2005}. However, this indicates that a fraction of quiescent green valley galaxies (20\%) have made this transition more than once, i.e. they have quiescent progenitors at higher redshifts, which transitioned back to the blue cloud or green-valley, and they are now on their way back to the red sequence.

\begin{figure*}[th]
\centering
\epsscale{1.12}
\plotone{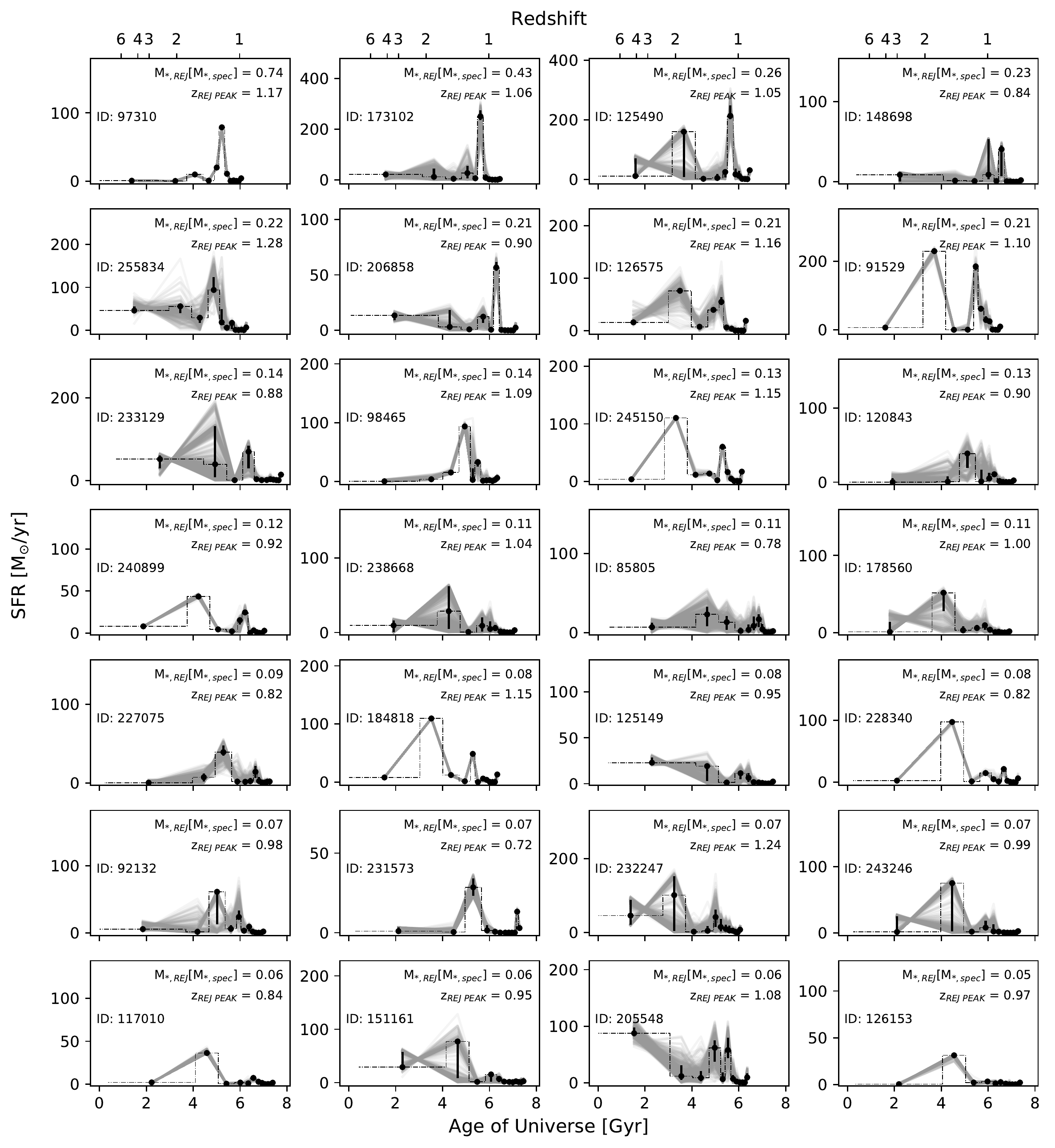}
\caption{The reconstructed star formation histories of rejuvenated galaxies obtained from MCMC full-spectrum fitting \citep[the walkers are shown in gray,][]{chauke2018}. The fraction of stellar mass formed from the rejuvenation episode and the redshift of the peak SFR of the event are shown in black.}
\label{fig:fit3}
\end{figure*}

\renewcommand{\thefigure}{\arabic{figure} (Continued)}
\addtocounter{figure}{-1}

\begin{figure*}[t]
\centering
\epsscale{1.12}
\plotone{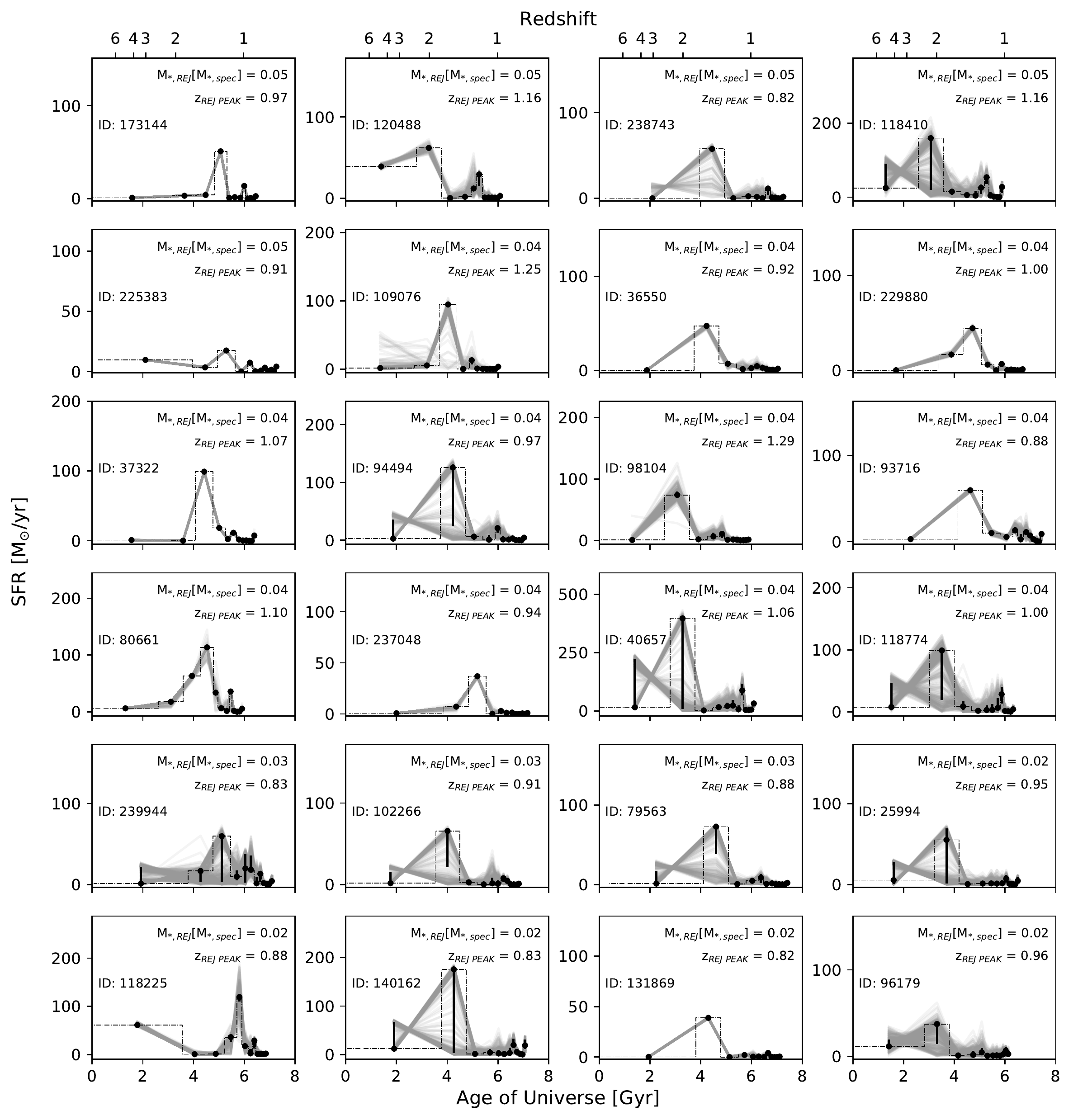}
\caption{}
\label{fig:fit4}
\end{figure*}

\renewcommand{\thefigure}{\arabic{figure}}

\subsection{SFHs and rejuvenation timescales}
\label{sec:time}

Figure \ref{fig:fit3} shows rejuvenated galaxies' reconstructed SFHs, the gray lines represent the MCMC walkers (see Section \ref{sec:datsfh}), the black points and the lower and upper error bars represent the $50^{th}$, $16^{th}$ and $84^{th}$ percentiles of the walkers, and the horizontal dashed lines show the sizes of the (constant star-formation) CSP age bins. The fraction of \mfit\ formed from rejuvenation as well as the redshift of the peak SFR of the episode are shown in black. The presence of young and old populations seen in the SFHs is driven by the presence, in the galaxy spectra (Figure \ref{fig:fit1}), of both features characteristic of young and older stellar populations, such as Balmer absorption lines and the G-band.

We find that galaxies in the LEGA-C sample have rejuvenation episodes during the redshift range $0.7<z<1.5$. This is because our method can only trace \replaced{secondary SF}{rejuvenation} events at lookback times $\lesssim10$Gyr because our oldest CSP bin is wide ($\sim3.5$ Gyr), therefore, the algorithm cannot trace rejuvenation for redshifts $z\gtrsim2$. However, it is not clear whether full spectrum fitting can trace rejuvenation with older stellar populations ($>5$ Gyr). On average, we find that galaxies that rejuvenate \deleted{(13\%)} first become quiescent at $z=1.2$ for about $\sim1$Gyr before their secondary SF episode, which lasts $\sim0.7$Gyr. \added{Non-rejuvenated galaxies in our sample first become quiescent, on average, at $z=1.3$. We note that the rejuvenation events we identify are distinct from stochastic variations in the SFR of galaxies on the star-forming sequence. The latter likely occur on shorter time scales ($\sim100$Myr) and are consequently averaged out in our SFH reconstruction.}

\begin{figure}[t]
\gridline{\fig{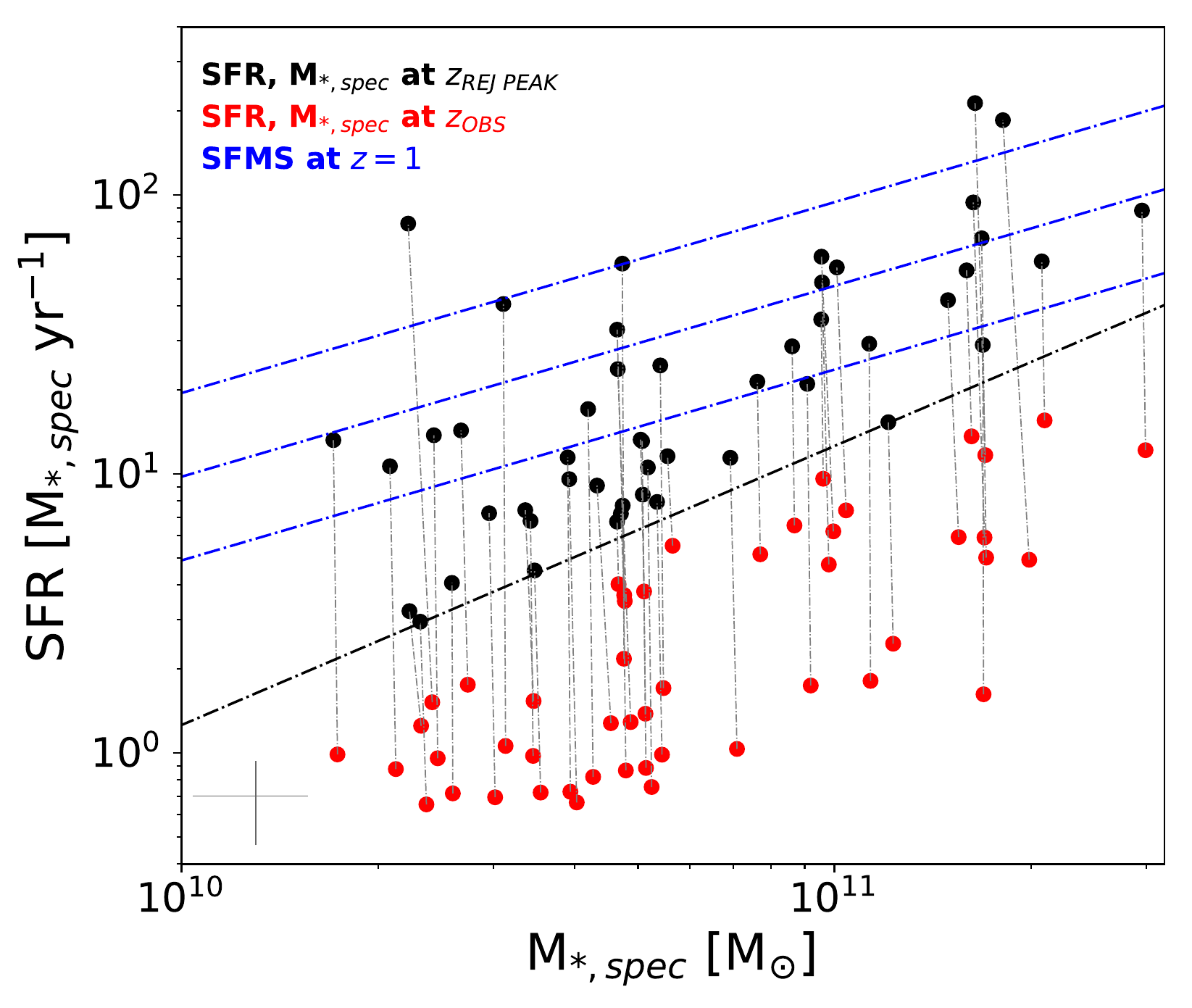}{0.45\textwidth}{}}
\caption{The peak SFR versus the stellar mass during the rejuvenation event (black) compared to the SFR-M$_*$ relation of the same sample at the redshift of observation (red). The gray connecting lines track the evolution of each galaxy in SFR and stellar mass. The blue dashed lines represent the \citet{speagle2014} SFMS at $z=1$ with a $0.3$dex scatter and the black dashed line distinguishes the star-forming sequence from the quiescent sequence. Typical error bars are indicated in dark gray.}
\label{fig:peaksfr}
\end{figure}

\begin{figure}[t]
\centering
\includegraphics[width=0.45\textwidth]{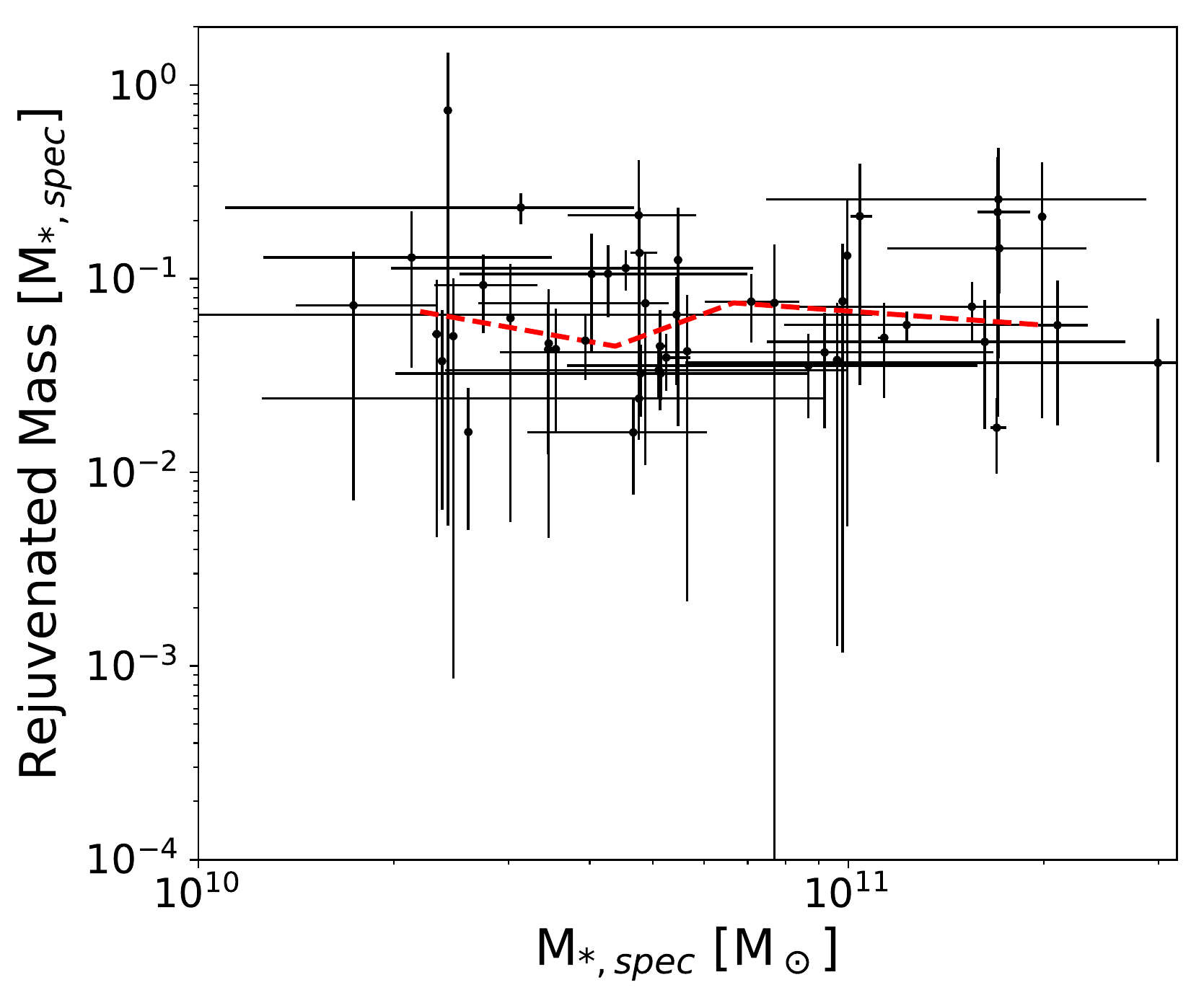}
\caption{Stellar mass of rejuvenated galaxies versus the fraction of stellar mass from the rejuvenation event. The upper and lower uncertainties are based on the 16$^{th}$ and 84$^{th}$ percentiles of the walkers (see Section \ref{sec:datsfh}). The median trend is indicated in red.}
\label{fig:pp}
\end{figure}

\subsection{SFR-Mass relation during rejuvenation}
\label{sec:dep}

In Figure  \ref{fig:peaksfr}, we show the SFR-M$_*$ relation of the rejuvenated sample at the peak of their SF episode (black) compared to the relation at their observed redshift (red) when they have transitioned back to quiescence. The gray connecting lines track the evolution of each galaxy in SFR and stellar mass, the blue dashed lines represent the star-forming main sequence (SFMS) at redshift $z=1$ \citep{speagle2014} with a $0.3$dex scatter, and the black dashed line distinguishes star-forming and quiescent populations at redshift $z=1$. This shows that rejuvenation episodes can lead to SFRs which are high enough to cause galaxies to transition back and forth between the SFMS (as well as the starburst region) and the red sequence. These galaxies cover a wide range of SFRs during their rejuvenation episodes, $\sim50\%$ are within 0.3dex of the SFMS and \replaced{2}{3} reach starburst status ($0.3$dex above the SFMS) during their rejuvenation episode. However, Figure \ref{fig:peaksfr} also shows that the stellar mass does not increase much after the rejuvenation episode (see Section \ref{sec:density}).

\added{
\subsection{Stellar mass and local environmental density dependence}
\label{sec:density}}

Figure \ref{fig:pp} shows the fraction of stellar mass attributed to rejuvenation as a function of \mfit. The median trend is indicated in red (computed using $\sim10$ galaxies per stellar mass bin) and the upper and lower uncertainties are based on the 16$^{th}$ and 84$^{th}$ percentiles of the walkers of the MCMC algorithm (see Section \ref{sec:datsfh}). On average, rejuvenation events result in the formation of a small fraction of stellar mass: they account for 10\% of the stellar mass of these galaxies, with 67\% of galaxies having rejuvenated masses $\leq0.1\times$\mfit. The median trend \deleted{indicates a weak anti-correlation between the stellar mass formed during rejuvenation events and stellar mass, although this trend is also} \added{is} consistent with a constant rejuvenation mass fraction with \replaced{galaxy}{stellar} mass. 

\begin{figure}[t]
\gridline{\fig{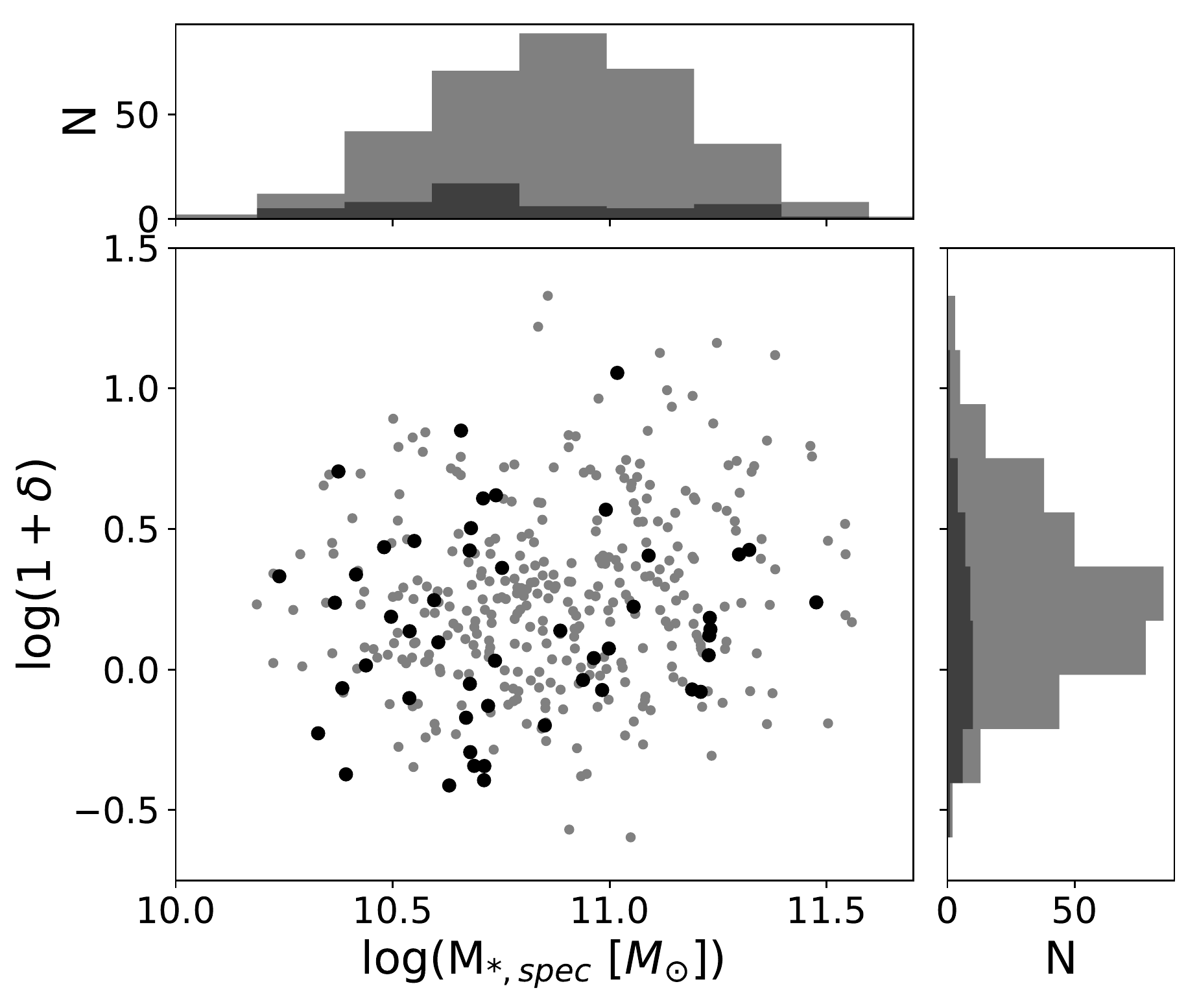}{0.45\textwidth}{}}
\caption{Local \replaced{density}{overdensity} versus the stellar mass of quiescent LEGA-C galaxies. The large black points represent galaxies that were rejuvenated. Distributions of the stellar mass and local \replaced{density}{overdensity} are shown on the top and right, respectively.}
\label{fig:dens}
\end{figure}

In Figure \ref{fig:dens}, we show the scale-independent local \added{environmental} density as a function of galaxy stellar mass. The gray points represent the quiescent sample and the black points represent rejuvenated galaxies. \added{The rejuvenated and non-rejuvenated populations span the same range in redshift and K-band magnitude and have similar distributions.} To test if rejuvenation is density dependent, we use a Kolmogorov-Smirnov test to compare the \replaced{density}{local overdensity} distribution of the rejuvenated sample to the overall quiescent sample. We find that the occurrence of rejuvenated galaxies increases with decreasing \added{local environmental} density (D statistic $=0.23$, $p<0.01$), which is in agreement with \citet{schawinski2007} (see Section \ref{sec:intro}). Rejuvenated galaxies have smaller stellar masses compared to the quiescent sample (D statistic $=0.36$, $p<0.01$). Most galaxies that show evidence of rejuvenation have stellar masses \mfit\,$<10^{11}$\mdot\ (see Figure \ref{fig:dens}). These trends are consistent with gas-rich mergers triggering rejuvenation events. Lower-mass galaxies in lower-density environments are more likely to merge with smaller gas-rich galaxies. On the other hand, in more dense environments, the gas in lower-mass galaxies is stripped by high-mass galaxies resulting in dry mergers.\deleted{ In contrast, we find no correlation between rejuvenation events and other galaxy parameters, such as size and Sersic index.}

\subsection{Contribution to the Cosmic Star-formation Rate Density}
\label{sec:sfrd}

To determine if \replaced{secondary SF}{rejuvenation} episodes contribute significantly to the stellar mass and SF budget in the universe, we compute\added{d} the fraction of stellar mass formed during rejuvenation events in quiescent galaxies \added{in Section \ref{sec:walk}}. \deleted{The detected rejuvenation events in our sample account for just $1$\% of the total stellar mass in quiescent galaxies at $z\sim0.8$, where we have applied a completeness correction on a galaxy-by-galaxy basis as described by \citet{wu2018} to create a volume-limited quantity. However, these rejuvenation events account for only $2\pm1$\% of the stellar mass in quiescent galaxies at $z\sim0.8$. The uncertainties are estimated by bootstrapping the sample.}\added{Rejuvenation events accounting for only $2\pm1$\% of the stellar mass in quiescent galaxies at $z\sim0.8$, together with the rejuvenation fraction, means that the}\deleted{The} average SFRD in the redshift range $0.7<z<1.5$ made up by rejuvenation events is $3\times10^{-4}$\mdot{yr}$^{-1}$Mpc$^{-3}$, a mere $0.3\%$ of the total \citet{madau2014} SFRD. This indicates that only a negligible fraction of all SF at this epoch is due to revived quiescent galaxies. 

\deleted{However, there will be rejuvenation events that are missed by the selection criteria described in Section \ref{sec:datsamp}, which motivates us to consider the probability that a rejuvenation event occurred in each quiescent galaxy. For this purpose we use the MCMC walkers to compute the probability that a galaxy is quiescent ($p_{Q_i}$) or star-forming ($p_{{SF}_{i}}$) for each of the 12 time bins $i$. We find the time interval where the galaxy is quiescent (i.e. $p_{Q_{i}}>0.5$), compute the maximum ($p_{Q_{qmax}}$), i.e. when the galaxy has the highest chance of being quiescent. $p_{Q_{qmax}}$ is multiplied to $p_{{SF}_{i}}$, where $i$ ranges from $qmax+1$ to 12 (the youngest bin), the total number of bins. We compute the maximum of $p_{Q_{qmax}}\cdot{}p_{{SF}_{i}}$ and then sum up the maxima in the quiescent sample to find the fraction of galaxies with secondary SF episodes. We also measure the mass formed between the first time a galaxy reached quiescence ($i=q1$) and $i=12$.}

\deleted{We find that $16\pm3$\% of the quiescent population has returned to the star-forming sequence during the epoch $z\sim0.7-1.5$ after reaching quiescence at some earlier time. This is consistent with the fraction of galaxies that have been identified as rejuvenated (Section \ref{sec:datsamp}).  However, these rejuvenation events account for only $2\pm1$\% of the stellar mass in quiescent galaxies at $z\sim0.8$. The uncertainties are estimated by bootstrapping the sample. These numbers are based on the requirement that a galaxy moves back to the SF sequence, as described in Sec \ref{sec:datsamp}.  If we repeat our calculation with a less strict requirement, namely that SF exceeds a fixed value of  log(sSFR [Gyr$^{-1}]) <-1$ at any redshift after initial quiescence, instead of an evolving limit, the fraction of galaxies with secondary SF events increases to $24\pm2$\%, with a total mass contribution of $4\pm1$\%.} 

\deleted{Our rejuvenation fraction lies between \citet{donas2007} and \citet{schawinski2007}'s measured values of 10\% and 30\%, respectively, for $z\sim0.1$ galaxies, though we note that our value is a lower limit  as we only trace secondary SF episodes in the redshift range $0.7<z<1.5$ due to limitations in the reconstructed SFHs (see Section \ref{sec:time}). Furthermore, \citet{donas2007} and \citet{schawinski2007} use UV-color relations to trace recent SF, therefore, their methods also trace secondary SF that does not necessarily lead galaxies back to the the star-forming sequence.}

\deleted{\citet{pandya2017} also measure a higher rejuvenation fraction (31\%) in their SAM (see Section \ref{sec:intro}). They define the quiescent region to be $1.4$dex below the SFMS, considerably more strict than our definition, and they additionally define a transition region ($0.6-1.4$dex below SFMS) between the star-forming and quiescent sequence. Defining our quiescent region in the same manner results in our rejuvenation fraction decreasing to 5\%. However, if we instead define the quiescent region to be $0.6$dex below the SFMS, i.e. combine \citet{pandya2017}'s transition and quiescent population, our rejuvenation fraction increases to 18\%, still lower than their value of 31\%. The discrepancy between these results may suggest that our sample contains galaxies with hidden rejuvenation at larger lookback times ($\gtrsim10$Gyrs), or that the fraction of rejuvenated galaxies is higher at $z=0$. However, we note that \citet{pandya2017}'s SAM underproduces quiescent galaxies at $z>0.5$, and they suggest that one of the reasons could be that quiescent galaxies are rejuvenating too much in their SAM.}

\section{Summary}
\label{sec:summ}

We have investigated \replaced{secondary SF}{rejuvenation} in quiescent galaxies in the LEGA-C sample using \citet{chauke2018}'s reconstructed SFHs, which were obtained from full spectrum fitting. We have shown that most galaxies which have had a rejuvenation episode lie in the green valley, i.e. they have intermediate U-V colors and stellar masses (Figure \ref{fig:uvm}). We presented the fraction of LEGA-C's quiescent population that have experienced rejuvenation events in their recent past, i.e. galaxies which had at one point transitioned from the star-forming to the quiescent sequence, transitioned back to the star-forming sequence before becoming quiescent again (Figure \ref{fig:fit3}). 

Limitations from the full spectrum fitting algorithm (see Section \ref{sec:disc}) means that we can only measure rejuvenation from redshift $z\lesssim2$ (lookback $\lesssim10$Gyr w.r.t. to the present day). We measure these events in the redshift range $0.7<z<1.5$ and they have an average time span of $\sim0.7$Gyr (Figure \ref{fig:fit3}). The occurrence of rejuvenated galaxies is higher in low-density environments, which is in agreement with previous studies (Figure \ref{fig:dens}). We found that rejuvenated galaxies generally have lower stellar masses ($<10^{11}$\mdot)  compared to the overall quiescent population, however, we do not measure a dependence on other galaxy parameters such as size.

On average, rejuvenation episodes generate 10\% of the galaxies' total stellar mass (Figure \ref{fig:pp}). At the peak SFR of the rejuvenation episode, many galaxies transition back to the SFMS (Figure \ref{fig:peaksfr}). $16\pm3$\% of the quiescent population has likely experienced a rejuvenation episode, this is only $2\pm1$\% of the stellar mass in the quiescent sample, which means that rejuvenation episodes in the redshift range $z\sim0.7-1.5$ account for only 0.3\% of the SFRD. This shows that although a significant portion of galaxies experience rejuvenation episodes, the mass formed from such events does not significantly contribute to the SFRD in the Universe. Therefore, rejuvenation is not an important factor in the growth of the red sequence, however, it can be a significant factor in detailed color studies.

\section{Acknowledgements}
\label{sec:ack}

Based on observations made with ESO Telescopes at the La Silla Paranal Observatory under programme ID 194-A.2005 (The LEGA-C Public Spectroscopy Survey).  PC gratefully acknowledge financial support through a DAAD-Stipendium. This project has received funding from the European Research Council (ERC) under the European Union's Horizon 2020 research and innovation programme (grant agreement No. 683184). KN and CS acknowledge support from the Deutsche Forschungsemeinschaft (GZ: WE 4755/4-1). We gratefully acknowledge the NWO Spinoza grant.


\bibliography{bibfile}

\appendix

\renewcommand{\thefigure}{\thesection.\arabic{figure}}

\section{Best-fit spectra of rejuvenated galaxies}
\label{app:a1}

\begin{figure}[h]
\centering
\epsscale{1.15}
\plotone{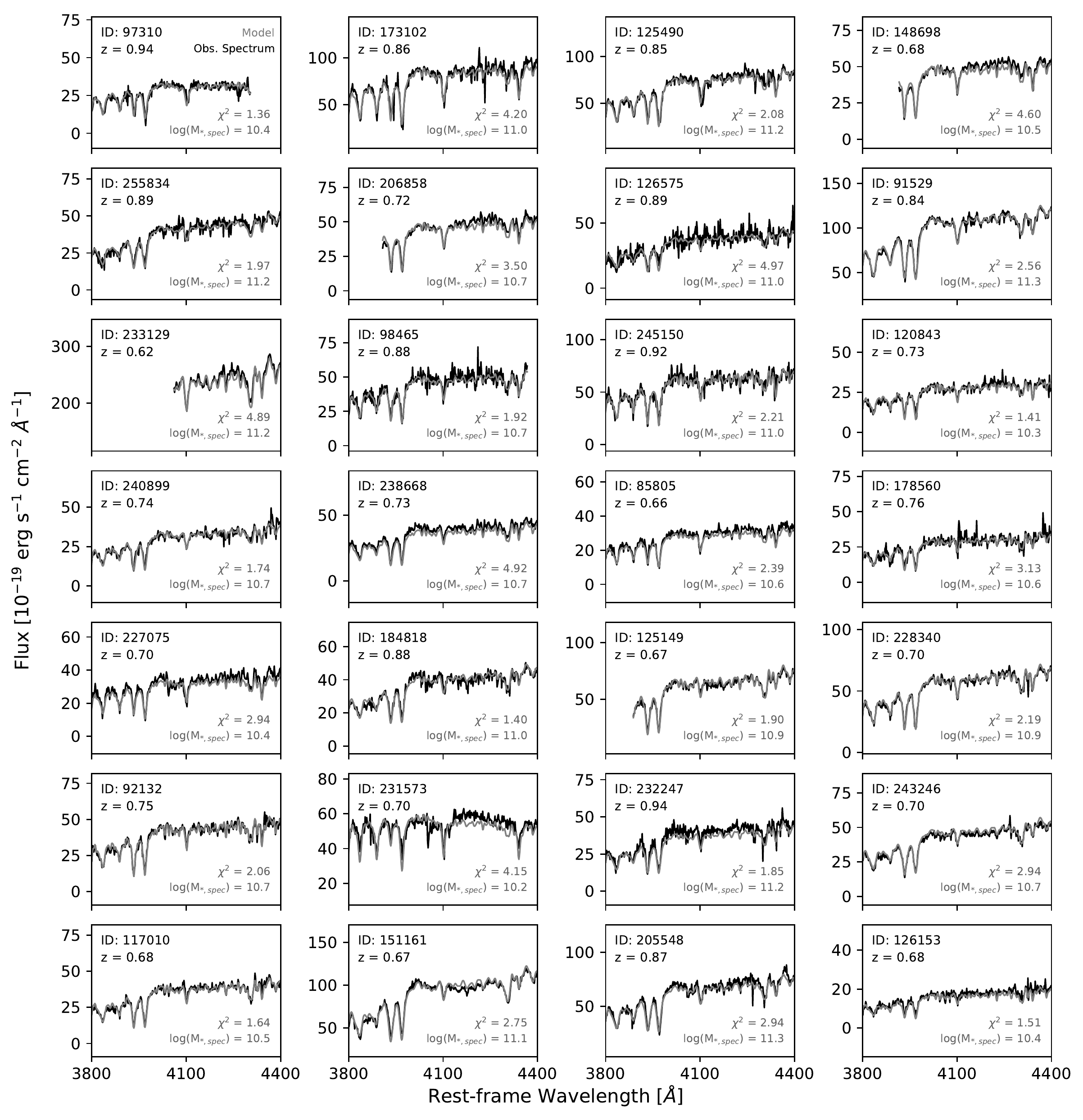}
\caption{Spectra of rejuvenated galaxies along with the resulting spectra obtained from MCMC full-spectrum fitting. Their IDs and redshifts are shown in black and the resultant normalised $\chi^2$ values and stellar masses are shown in gray.}
\label{fig:fit1}
\end{figure}

In Figure \ref{fig:fit3} we showed the reconstructed SFHs, obtained from full-spectrum fitting \citep{chauke2018}, of rejuvenated galaxies in the LEGA-C Survey. In Figure \ref{fig:fit1} we show the spectra of those rejuvenated galaxies, along with the resulting spectra obtained from full-spectrum fitting.

\renewcommand{\thefigure}{\thesection.\arabic{figure} (Continued)}
\addtocounter{figure}{-1}

\begin{figure}[h]
\centering
\epsscale{1.15}
\plotone{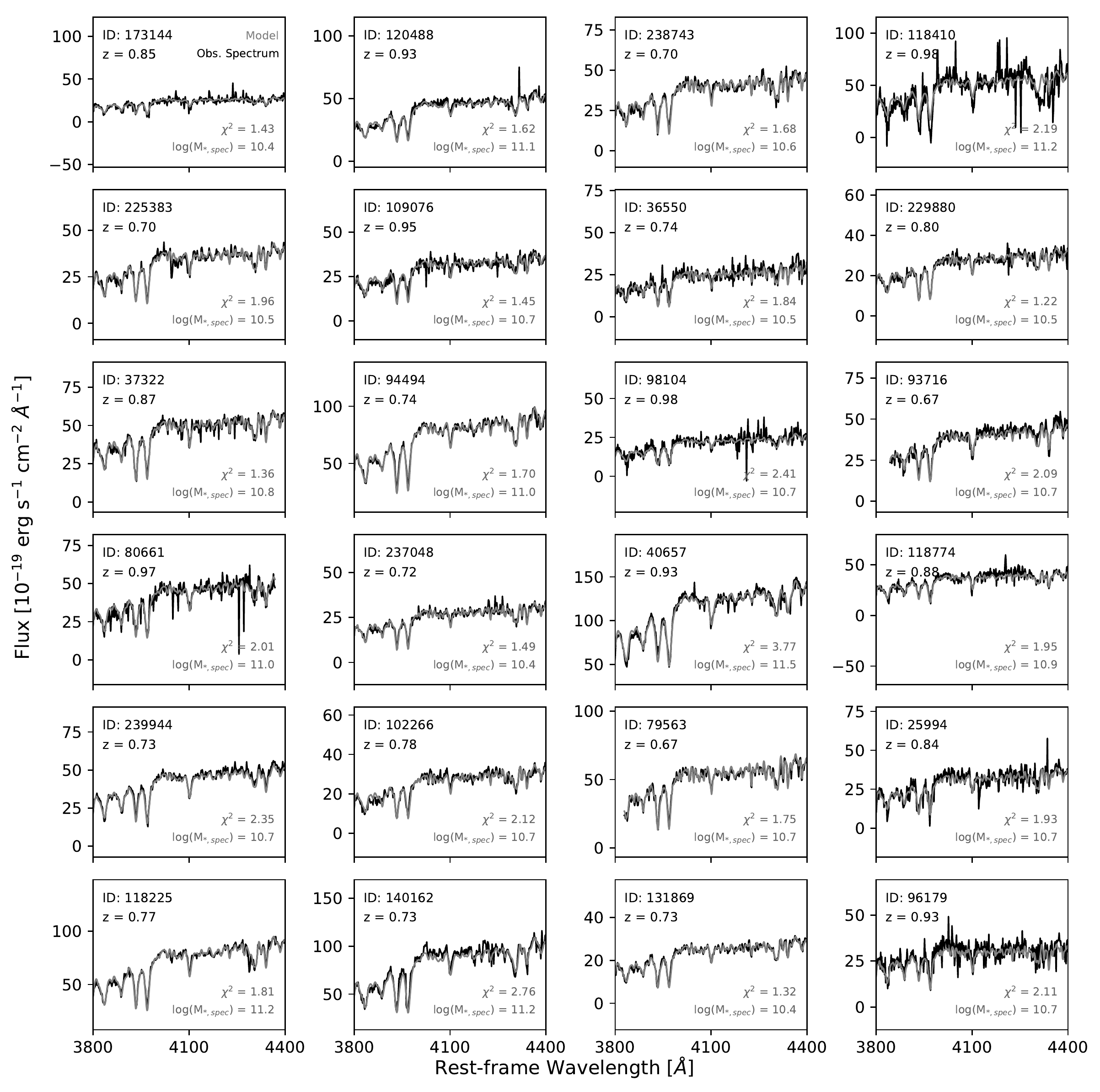}
\caption{}
\label{fig:fit2}
\end{figure}

\renewcommand{\thefigure}{\arabic{figure}}

\listofchanges

\end{document}